\documentclass[aps,pra,superscriptaddress,showpacs,floatfix
,twocolumn
]{revtex4}
\usepackage{epsf,epsfig,amsmath,mathrsfs}
\newcommand{\eq}[1]{Eq.~(\ref{#1})}
\newcommand{\be}[1]{\begin{equation}\label{#1}}
\newcommand{\fig}[1]{Fig.~\ref{#1}}
\newcommand{\figs}[3]{Figs.~\ref{#1},\ref{#2}, and \ref{#3}}
\def\hp{He$^+$}
\def\PRA{Phys. Rev. A\ }

\def\PRL{Phys. Rev. Lett.\ }
\begin{document}

\title{Intermanifold similarities in partial photoionization cross sections of helium}
\author{Tobias Schneider}
\affiliation{Max Planck Institute for the Physics of Complex Systems, N\"othnitzer Str. 38, D-01187 Dresden, Germany}
\author{Chien-Nan Liu}
\affiliation{Department of Physics, Cardwell Hall, Kansas State University, Manhattan, Kansas 66506, USA}
\author{Jan-Michael Rost}
\affiliation{Max Planck Institute for the Physics of Complex Systems, N\"othnitzer Str. 38, D-01187 Dresden, Germany}

\date{\today}

\begin{abstract}
Using the eigenchannel R-matrix method we calculate partial
photoionization cross sections from the ground state of the helium
atom for incident photon energies up to the $N=9$ manifold. The wide
energy range covered by our calculations permits a thorough
investigation of general patterns in the cross sections which were
first discussed by Menzel and co-workers [Phys. Rev. A {\bf 54}, 2080
(1996)]. The existence of these patterns can easily be understood in
terms of propensity rules for autoionization. As the photon energy is
increased the regular patterns are locally interrupted by perturber states
until they fade out indicating the progressive break-down of
the propensity rules and the underlying approximate quantum numbers.
We demonstrate that  the destructive influence of
isolated perturbers can be compensated with
 an energy-dependent quantum defect.
\end{abstract}
\maketitle
%
\section{Introduction}
Consisting of  only three particles, two electrons and a
nucleus, the helium atom  nevertheless   possesses rich dynamics with
 complex features. Hence,  helium  has  always been a focus of research
and it has incessantly been used as a testing
ground of fundamental concepts. In every energy regime the correlated
dynamics of the two electrons can be probed by photon impact. This
has been the most precise method of investigation in terms of energy
resolution although some limitations exist since only those excited states can
be accessed whose population is allowed via dipole
selection rules from the initial state (usually the ground state).
The domain of high double excitation can be scrutinized
in greater detail with the advance of the experimental and
theoretical tools available.  A leading theme in these studies is the
exploration of regularities in the observables of this classically
chaotic three-body Coulomb system.
Moreover, one would like to know how chaotic features emerge when the
double-ionization threshold is approached by increasing the energy.

Most experiments have concentrated on total photoabsorption cross
sections \cite{MaCo63,MoEd84,KKV88,Doal}. This is also true for those
theoretical calculations which have reached the highest excitation
energy so far.  The reason is simply that the method of complex
rotation allows for the most effective computation of resonances in
terms of their complex energies (where the real part is the energy
position and the imaginary part half the resonance width).  However,
these widths are total widths and only non-differential observables
such as the total photoabsorption cross section can be constructed
without losing the effectivity of the approach
\cite{WiDe93,Roal97,GrDe98,Bual95}.

Yet, as has been demonstrated recently, interesting additional
features such as radiative and relativistic effects emerge by, {\it
e.g.}, measuring the photon emission following the photo excitation \cite{Rual,Odal00,Peal01,LCL01}.
It turns out that this signal reveals the splitting of the \hp\
threshold due to spin-orbit coupling.
Moreover, in some experiments \cite{WuSa82,Zual89}  partial photoionization
cross sections following photoabsorption into doubly excited states
have been measured, most recently up to energies of the $N=5$ excitation threshold
of \hp\ \cite{Meal}.
In the latter work similarities between  partial cross
sections belonging to different manifolds
have been observed and related to the propensity rules for doubly
excited states \cite{RoBr90,SaGr90,TRRreview}. This type of similarity
has to be distinguished from {\it intra}manifold similarities of
partial photo cross sections such as mirroring and mimicking, as first
noted by Liu and Starace \cite{LiSt}.

In the present work we explore the origin of {\it inter}manifold
similarities of partial photoionization cross sections in detail.  As
a function of increasing excitation energy we will describe and
explain how these similarities emerge and begin to disappear again for
very high excitation.  To this end, we have calculated the partial
ionization cross sections up to the ninth threshold of \hp.  The
corresponding energies are much higher than those which were reached
previously, experimentally as well as numerically.  This allows us to
work out the similarities of the cross section pattern across eight
manifolds and to illustrate in detail how the propensity rules lead to
those similarities.  Our results confirm Menzel's conclusions for the
energy regime he considered.  Perturber states, emerging as a new
feature at higher excitation energies, seem to destroy the similarity
pattern.  However, as we will show, a regularization based on an
energy-dependent quantum defect can be introduced which restores the
similarities even in the presence of isolated perturbers.  The paper
is organized as follows: In section II we present the partial cross
sections and briefly describe computational details.  In section III
we briefly summarize the propensity rules for dipole excitation and
for autoionization of two-electron resonances, as well as their
classification schemes.  We also recall adiabatic two-electron
potential curves which facilitate the understanding of the
classification and propensity rules, before we formulate the general
scheme of the intermanifold patterns with the help of the propensity
rules.  In section IV we interpret the patterns of the partial cross
sections across the manifolds with this scheme.  The paper ends with a
summary in section V.

\section{Partial photoionization cross sections of helium up to the N
=9 level of \hp}

\subsection{Numerical procedure}

In the present study, the eigenchannel R-matrix method
\cite{rmat1,rmat2} combined with a close-coupling scheme
\cite{pan} is employed in order to calculate the partial cross
 sections for single photoionization of the helium atom. 

The eigenchannel R-matrix method has been successfully
applied to single photoionization \cite{chg} and photodetachment
\cite{he-} of atomic systems with two active electrons. The most
important concept of the R-matrix theory is to partition the
configuration space into two regions, namely the reaction region, where the
short-range interactions between one particle and a compact target
are complicated, and the external region, where the system can be
reduced to a two-body problem involving long-range interactions. 
For the current study, the reaction region is that part of
six-dimensional configuration space for which both electron lie within a
sphere of radius $r_0$. The reaction surface $\Sigma$ is the set of
points for which max$(r_1,r_2)=r_0$, where $r_1$ and $r_2$ are the
electron distances from the nucleus. The
method has been described in detail in the literature \cite{rmat2,pan,lerouzo}. Therefore, we
present here only a brief overview and some numerical details.

Within the reaction region, using a set of Slater-determinants composed of properly chosen
one-electron orbitals, the electron-electron interaction is
fully taken into account by applying bound-state configuration
interaction techniques. At a given energy the eigenchannel R-matrix
method aims to determine varationally a basis set of wavefunctions, the
so-called R-matrix eigenchannel wavefunctions, which are orthogonal and
complete over the reaction surface $\Sigma$ enclosing the reaction
region, and their negative logarithmic derivatives being constant over
$\Sigma$. 
The helium wavefunctions of experimentally observed channels
can be represented by linear combinations of the eigenchannel
wavefunctions thus constructed within the reaction region.

In the external region, since only single ionization processes are
considered, it is assumed that there is only a single electron while the other
electron is bound. Instead of applying the conventional multichannel
quantum defect theory \cite{rmat2}, Pan {\it et al.} \cite{pan} developed an approach using a
close-coupling scheme to obtain a basis set of multichannel wavefunctions which describe
the outgoing electron and the residual ion. In addition to the Coulomb potential, all
multipole interactions in the external region are included numerically to
account for the polarization of the residual ion. Note that, although
the asymptotic behavior of a one-electron continuum wavefunction in
a Coulomb field is well known \cite{smith}, this description of a 
singly ionized state in a two-electron atom it is exact only at an
infinite distance from the nucleus. Since one can only integrate the
close-coupling equation starting from a finite distance,
we use WKB representations \cite{wkb} for the wavefunction instead at
a suitably large distance. To describe an experimentally observed
channel, one has to form a linear combination of these multichannel
basis wavefunctions according to the incoming-wave boundary
condition \cite{pan}.

By matching the linear combinations of the multichannel
basis functions for the two regions, one can determine the exact
final state wavefunctions $\Psi{_i}^{(-)}$ which describe the experimentally observed
channels $i$. The partial cross sections can be calculated according to the
standard formula \cite{starace}:
\begin{equation}
\sigma_i=\frac{4\pi^2\omega}{c}|\langle\Psi{_i}^{(-)}|D|\Psi_0\rangle|^2,
\label{cross_section}
\end{equation}
where $\omega$ is the photon energy, $D$ is the dipole operator, and $c$ is the speed of light. The wavefunction $\Psi_0$ in \eq{cross_section} denotes the helium ground state.

In the present study, the radius $r_0$ of the R-matrix sphere is
chosen to be 200 a.u.. A total of 1080 closed-type ({\it i.e.}, zero
at the radius $r_0$) and 20 open-type ({\it i.e.}, non-zero at the
radius $r_0$) one-electron wavefunctions with orbital
angular momentum up to 9 are included. 9610 closed-type two-electron
configurations are included in the calculation for the final state
wavefunction. For each channel in which one electron can escape from
the reaction region, two open-type  orbitals for the outer electron
are included in addition to the closed-type basis set. For a given
photon energy, besides all open channels, relevant closed channels
are also included in the calculations (cf. \cite{pan}).

\begin{figure}
\epsfxsize  \columnwidth
\centering\epsffile{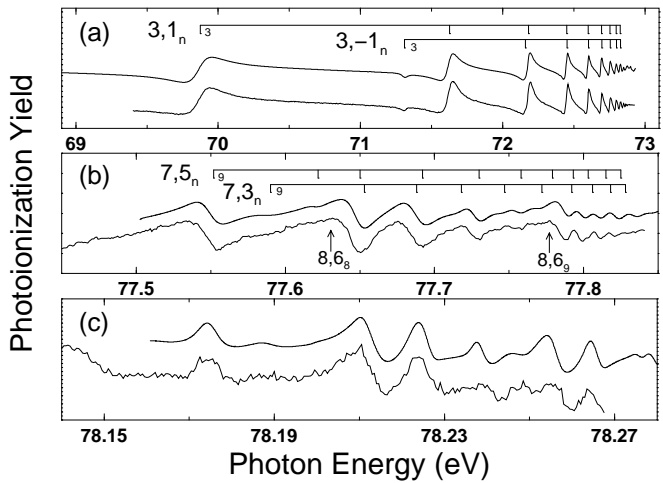}
\caption[]{Calculated total photoionization cross section in
comparison with experimental results of Refs.~\onlinecite{Doal,Pu99}
below the thresholds (a) $N=3$, (b) $N=7$, and (c) $N=9$.  The
theoretical data (thick lines) has been shifted upwards to allow for
an easier comparison with the experiment (thin lines).  Moreover, the
numerical data has been convoluted with a Gaussian of 5\ meV width for
$N=3,7$ and 2\ meV for $N=9$.  In (a) and (b) the positions of the
resonance states of the two strongest Rydberg series are indicated
\cite{Roal97}, for an explanation of the quantum numbers, see section 
III.}
\label{totalcs}
\end{figure}
\begin{figure}[b]
\epsfxsize  .95\columnwidth
\centering\epsffile{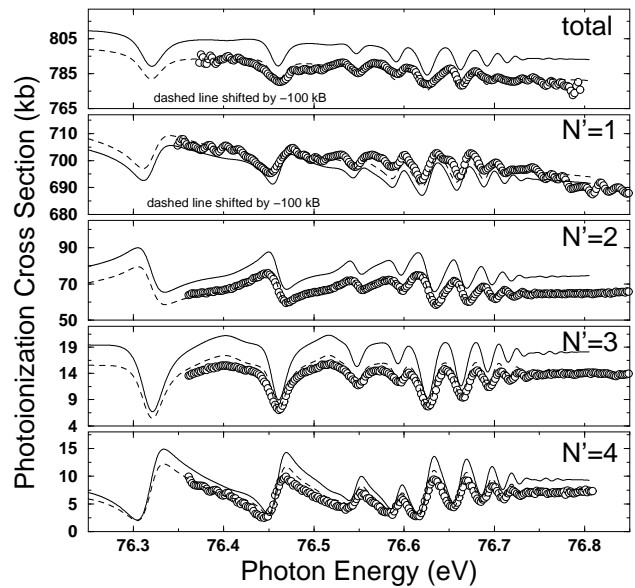}
\caption[]{Calculated (absolute) total and partial cross sections are
compared to experimental data (circles) of Menzel {\it et al.}
\cite{Meal} in the region of the $N=5$ resonances.  Calculation in
velocity gauge: solid lines; calculation in acceleration gauge: dashed
lines; experiment: open circles.  The numerical results are shown for
the velocity gauge (solid) and the acceleration gauge (dashed) and
have been convoluted with a Gaussian of 5\ meV width.  The
acceleration gauge result for $N'=1$ (and consequently for the total
cross section) is shifted by $-100$ kb.}
\label{N5partials}
\end{figure}
\subsection{Typical cross sections}
Our calculated total photo cross section below the $N=3,7$, and $9$
threshold are shown in \fig{totalcs} together with experimental data
by Kaindl and co-workers \cite{Doal,Pu99}.  Since no absolute
photoionization yields are measured in these experiments we have
scaled the experimental data to our results.  As can be clearly seen
the calculated cross section is in excellent agreement with the
experimental one.

\fig{N5partials} shows the total and partial cross sections below the
$N=5$ threshold of \hp.  
Note that below a given threshold $N$ we are dealing with $N-1$
partial cross sections $\sigma_{N,N'}\ (N'=1,\ldots,N-1)$, where $N'(<N)$ denotes the principal quantum number of the residual
helium ion. Hence, in the case of the $N=5$ threshold we are concerned with four partial cross sections, namely, $\sigma_{5,1}\ldots\sigma_{5,4}$. 
The agreement with existing experimental data
\cite{Meal} on an absolute scale is in general good.  Interestingly
the cross section in acceleration gauge (dashed) is higher than in
velocity gauge (solid) and higher than the experiment for the partial
cross section $N'=1$.  For all other partial cross sections the
velocity gauge result is too high and the acceleration gauge matches
the experiment better.  The total cross section behaves as the $N'=1$
partial cross section by which it is dominated.  This observation of
numerical accuracy points to a fundamental difference of the $N'=1$
cross section compared to all other partial cross sections which is
also confirmed by the fact that $N'=1$ takes about 90\% of the yield
while the yield for the higher partial cross sections decreases with
increasing $N'$ but only slightly.

Since we focus on the general patterns of the partial cross sections
which agree in both gauges very well with experiment the minor
discrepancies in the absolute value are of no concern.

\subsubsection*{Partial cross sections of the resonances converging to
the $N=9$ threshold of \hp}

\begin{figure}[t]
\epsfxsize  \columnwidth
\epsffile{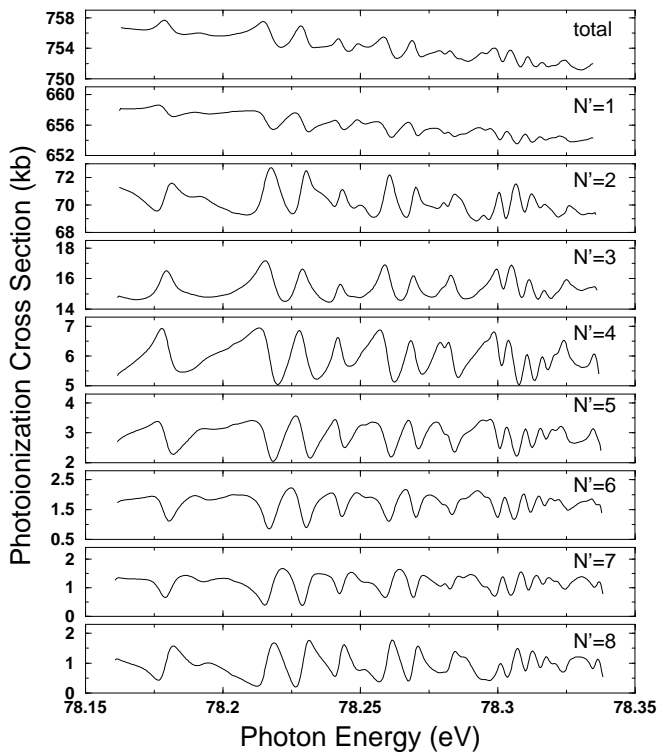}
\caption[]{Calculated partial photoionization cross sections below
the $N=9$ threshold (using velocity gauge). The data is convoluted with
a Gaussian of  1\ meV width.}
\label{N9partials}
\end{figure}

The ninth threshold is only about 0.67 eV below the double ionization
threshold. Partial cross sections for the
 $N=9$ manifold have neither been measured  nor been  calculated so far.
 Based on the good agreement of the total
cross section with the experiment (see \fig{totalcs}) we believe that our
calculation in this energy range is still reliable. As can be seen
in \fig{N9partials} the regular sequences of Rydberg series observed
for lower manifolds appear to be lost. However, even if a regular
Rydberg progression exists it is very difficult to identify it at a
finite energy resolution since the peaks accumulate
towards threshold. For this reason we will use an alternative way to
represent the cross section data.

\subsubsection*{Unfolding cross sections}
\label{subsubsec:scal}
\begin{figure}[b]
\epsfxsize  \columnwidth
\epsffile{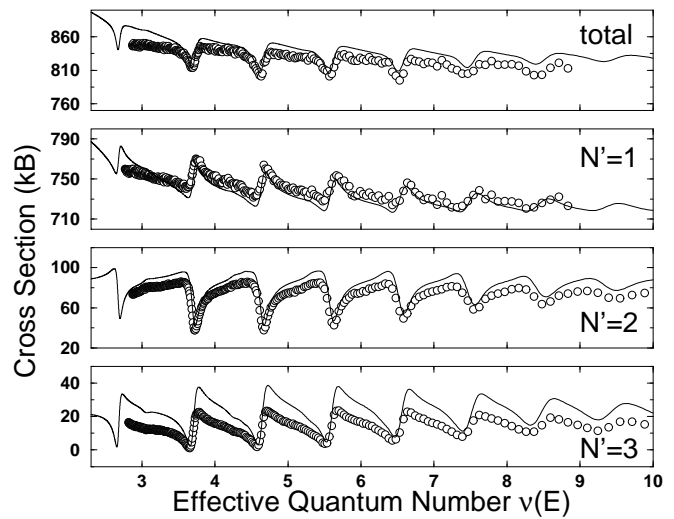}
\caption[]{Partial cross sections as a function of the effective
quantum number $\nu_4(E)$ below the $N=4$ threshold. Due to an
energy-independent quantum defect the resonance spacings are equal.
Theory (velocity gauge): solid lines; experiment\cite{Meal}: open
circles. The numerical data is convoluted with a Gaussian of width 5\
meV.}
\label{N4partials}
\end{figure}

To make all peaks of  a Rydberg progression in a cross section clearly
visible which is particularly important for analyzing similarities in
the patterns of cross sections we re-parametrize the energy according
to the effective quantum number. An ideal unperturbed Rydberg series
converging to a threshold $N$ of the He$^+$ ion would have
equidistant peaks  as a function of the effective quantum number
\begin{equation}
\nu_N(E)=\sqrt{\frac {{\cal R}}{I_N-E}},
\label{scaling}
\end{equation}
where ${\cal R}$ is the Rydberg constant, and $I_N=4{\cal R}/N^2$
denotes the $N$th ionization potential ($I_\infty=0 {\rm\ a.u}.$) of
He$^+$. In
\fig{N4partials} we show partial cross sections below the $N=4$
threshold where we have scaled the energy axis according to \eq{scaling}.
The constant spacing of the resonances indicates  unperturbed Rydberg
series. Note also that the two
partial cross sections \hp$(N'=1)$ and \hp$(N'=3)$ in \fig{N4partials}
behave quite similarly while \hp$(N'=2)$ mirrors their pattern. This
mirroring and mimicking behavior of partial photo cross section is a
universal  {\it intra}manifold feature \cite{LiSt}.

So far, we have presented illustrative examples for the cross sections
to highlight the accuracy of our calculation.  We will systematically
present the {\it inter}manifold similarities between certain partial
cross sections and discuss their origin in the next two sections.  The
relations between certain chains of partial cross sections as well as
the interpretation which resonances contribute to them is based on the
existence of approximate quantum numbers and propensity rules for the
resonances which we will discuss first.

\section{Approximate quantum numbers and propensity rules}
\label{sec:prop}
\begin{figure}
\epsfxsize  \columnwidth
\epsffile{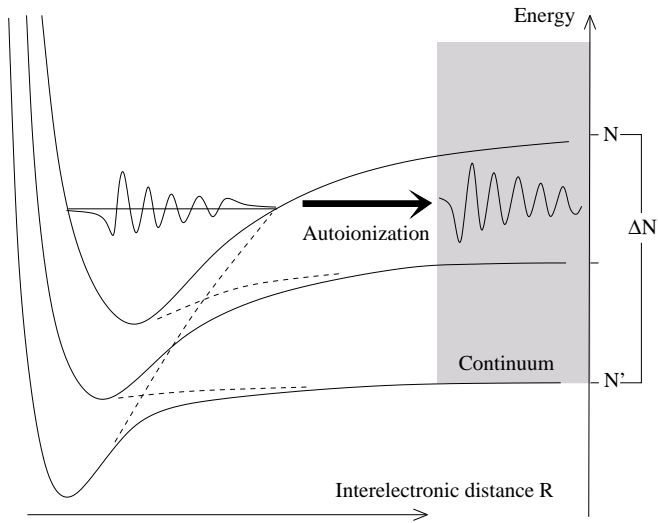}
\caption[]{Schematic representation of adiabatic potential curves. In
the adiabatic picture the resonances appear as vibrational
eigenstates. The mechanism of autoionization relies on non-adiabatic
transitions in this description. The dashed lines indicate the
avoided crossings of the potential curves which play an important
role in the derivation of the propensity rules (see Ref.~
\onlinecite{Roal97}).}
\label{adiacurve}
\end{figure}

Over the last 20 years a scheme of approximate quantum numbers for
doubly excited states has been developed which reflects the correlated
two-electron dynamics.  They have been expressed as $_{N}(K,T)^A$ by
Herrick and co-workers \cite{Heal} and assigned to hyperspherical
potential curves by Lin \cite{Lin}.  Feagin and Briggs \cite{FeBr}
gave a justification for the quantum numbers in terms of constants of
motion for a separable Hamiltonian which arose from the introduction
of the so-called molecular adiabatic approximation.  This
approximation is similar to the Born-Oppenheimer approximation for a
diatomic molecule, namely $H_2^+$, but with reversed roles of
electrons and nuclei.  In two-electron atoms it is the interelectronic
axis $R$ which is taken as adiabatic, {\it i.e.}, slow variable in
analogy to the internuclear axis in $H_2^+$.  In this picture the
doubly excited states naturally appear as vibrational eigenstates in
the adiabatic potential curves (cf.  \fig{adiacurve}).

The probably most simple way of understanding the quantum numbers is
to interpret them as the Stark quantum numbers of the inner electron
whose Coulomb motion is perturbed by the electric field of the outer
electron.  The quantum numbers remain the same along a Rydberg series
when the outer electron's quantum number $n$ increases to infinity
(single ionization limit) and the inner electron remains in the $N$th
excited state of the ion, where $N = N_1+N_2+m+1$ is the sum of the
Stark quantum numbers.  Note, that the classification will be relevant
to understand the pattern in partial ionization cross sections since
it is also applicable to singly ionized two electron states ({\it
i.e.}, continuum states).  The Stark quantum numbers (often called
parabolic quantum numbers) are related to Herrick's scheme by $T = m$
and $K = N_{2}-N_{1}$.  The label $A$ denotes the symmetry with
respect to the line $r_{1}=r_{2}$ in the wavefunction where the
$r_{i}$ are the electron-nucleus distances. The
complete signature of a two-electron resonance is then
$[N_{1}\,N_{2}\,m]_{n}^A$ or $_{N}(K,T)^A_{n}$.  For the
classification of resonance states in helium photoionization from the
ground state one very often uses a simplified Herrick's notation,
namely, $N,K_n$, where the other quantum numbers are redundant due to
the dipole selection rules (see Fig.~\ref{totalcs}).  For a more complete comparison between
the different quantum numbers see \cite{Roal97,TRRreview}.

The approximate constants of motion for correlated two-electron
dynamics expressed through the approximate quantum numbers imply a
nodal structure for the respective resonance states \cite{RBF91}. In
turn this nodal structure leads to preferences for
autoionization \cite{RoBr90} and (radiative) dipole transitions
\cite{VRB91}.

\subsection{Propensity rules for radiative transitions}
\label{subsec:radiative}
Propensity rules for radiative transitions can be derived by
analyzing the dipole matrix elements according to the  nodal
structure of the resonance wavefunctions,  which is a simple
analytical task on the potential saddle for $2\vec r \equiv \vec r_1
+ \vec r_2=0$. This region in configuration space is most relevant
for symmetrically excited electrons with $N\approx n$. It corresponds
to the equilibrium geometry of a linear $ABA$ molecule \cite{HuBe87}.
Not surprisingly, the relevant quantum number
\begin{equation}
v_2 = 2N_1+m
\label{v2}
\end{equation}
for radiative propensities quantizes  the two-fold  degenerate
bending motion of triatomic molecules and can be derived by normal
mode analysis about the saddle point \cite{VRB91,RoBr91}. Dipole
matrix elements within the saddle approximation follow the selection
rule
\begin{equation}
\Delta v_2 = 0,\pm 1
\label{dipo}
\end{equation}
that survives for the full dynamics as a propensity rule. Here, we
are interested in photoabsorption into doubly excited
states from the ground state of helium.  The final $A = +1$ states
with the admixture of lower channels for the relatively best overlap
with the ground state can only be $m= +1$ states due to the $^1\!P^o$
symmetry. Therefore, we expect a preference for
\begin{equation}
\Delta v_2 =1
\label{dipf}
\end{equation}
transitions ({\it i.e.,} $\Delta m = 1$).  In each manifold $N$ there
is only one series $[0\,(N-2)\,1]^+$ fulfilling this condition. This
series is commonly referred to as the {\it principal series} in the
literature. Other series (with $A = +1$) are also populated without
the preference of $\Delta N_{1}=0$.  However, they carry much less
oscillator strength.

\subsection{Propensity rules for non-radiative transitions}
The mechanism of autoionization relies on non-adiabatic transitions
in the (molecular) adiabatic picture. The rules for autoionization
can be stated by establishing a preference for nodal changes in the
wavefunction.

Most easily, $N_{2}$ can be changed which is the preferred decay mode.
This is achieved in the molecular description (as well as in the
hyperspherical one) by so called radial coupling matrix elements which
are large between states which differ only in $N_{2}$.  Rotational
coupling is only slightly less effective and changes the quantum
number $m$.  Finally, there is no mechanism to change $N_{1}$.  Hence,
a resonance decays only through changing $N_{1}$ if no other
possibility exists.

In parallel, the symmetry $A$ plays an important role.  In general,
states with $A=+1$ decay more easily than states with $A = -1$ which
can be seen from the narrower avoided crossings for $A=+1$ leading to
larger radial couplings compared to $A=-1$ states.
We may summarize the propensity rules for autoionization
\cite{RoBr90} according to the relative efficiency
of the underlying decay mechanism:
\begin{subequations}
 \label{nrprop}
 \begin{eqnarray}
        &\mbox{(A)\hspace{.5cm}}&{\rm reduction\ of\ }N_2
        \label{propAA} \\
        &\mbox{(B)\hspace{.5cm}}&{\rm change\ of\ } m
        \label{propAB} \\
        &\mbox{(C)\hspace{.5cm}}&{\rm reduction\ of\ } N_{1}.
        \label{propAC}
 \end{eqnarray}
\end{subequations}

These propensity rules group the $^1\!P^o$ resonant states of helium
into three classes I-III with typical widths separated by at least
two orders of magnitude, $\Gamma_{\rm I} : \Gamma_{{\rm II}} : \Gamma_{{\rm III}}
\approx 10^4 : 10^2 : 1$.
Since the propensities depend on the nodal structure $[N_1\,N_2\,m]$
of the inner electron, they  hold for entire Rydberg series
(different $n$) characterized by a single $[N_1\,N_2\,m]$
configuration. III.~class states for $^1\!P^o$ resonances
are restricted to the $[(N-1)\,0\,0]^-$ configurations, which enforce
decay through a $\Delta N_1 \ne 0$ transition (C).

\subsection{Propensities for partial photoionization cross sections}
We proceed now to formulate the conditions for the similarity in 
patterns of partial photo cross sections based on the exisiting 
propensity rules.
  The propensity rule (A) characterizes by far the
most important mechanism for autoionization and it is 
 this decay mechanism which also  determines the similarity patterns.

\subsubsection{Configurations, manifolds and chains}
\label{subsubsec:def}
So far we have already used the terms configuration and manifold.
A {\it configuration}  is a set of two-electron states
characterized by the quantum numbers $[N_1\,N_2\,m]^A$ which refer to the
state of the inner electron in the correlated two-electron state.
A {\it manifold} $N$  of two-electron states contains all
configurations whose quantum numbers add up to $N = N_{1}+N_{2}+m+1$.
Physically, one can think of $N$ being the principal quantum number
of the  electron in the He$^+$ ion which would remain  if the
outer electron would be taken away. In the adiabatic picture a configuration is represented
by a potential curve (see \fig{adiacurve}). This  illustrates that
the actual state of the outer electron is not specified for a
configuration. It can  be a bound state with quantum number $n$
in the potential curve corresponding to a resonance for the two-electron system.
This type of states we call in the present context {\it intermediate
configuration}.
The outer electron can also be in the continuum characterized by the
potential of the configuration of the inner electron. This type of
states we call {\it final configuration}.
It is important to realize that all the propensity rules refer to the
nodal character induced by the configurations only, {\it i.e.} to a first
approximation, the state of the
second electron being a Rydberg electron $n$ or in the continuum is
irrelevant for these propensities. However, this does not mean that
we deal with independent electron states. Rather the Stark quantum
numbers of a configuration for the inner electron characterize a whole
set of correlated two electron states.

Photoionization proceeds from the initial state (here the ground state
of helium) either directly or via a resonance of the intermediate
configuration to the final configuration.  Particularly the partial
cross sections $\sigma_{N,1}$ have a strong direct channel (simply the
photoionization of one of the electron).  This is seen in the large
smooth background cross section ({\it e.g.}, \fig{N5partials}) for the
$N'=1$ partial cross sections.

To form a partial cross section $\sigma_{N,N'}$ one has to take into
account all accessible intermediate and final configurations.  The
propensity rules can be used to structure the contributions of
different configurations and they determine which of these
configurations contribute dominantly to the cross section.
\begin{figure}
\epsfxsize  .8\columnwidth
\epsffile{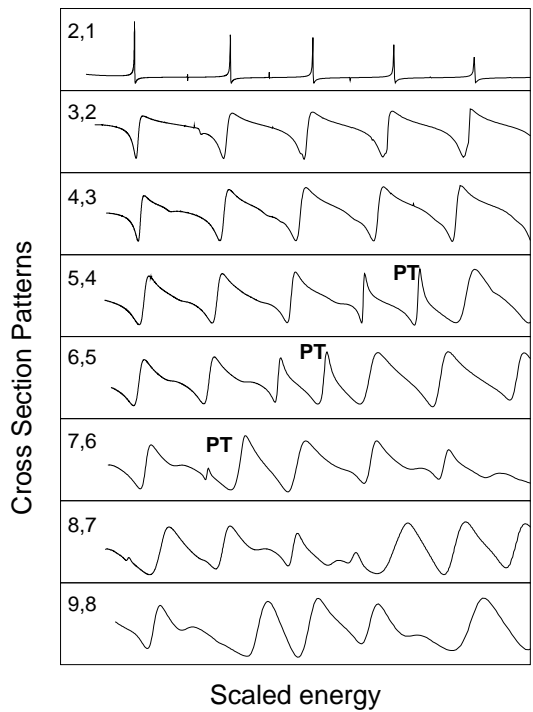}
\caption[]{Partial cross sections with $\Delta N=N-N'=1$. The energy
axis of each panel is scaled according to \eq{scaling}. Additionally, 
each of the individual cross sections is 
horizontally shifted by a constant quantum defect in order to 
approximately align the resonances of the principal series. The numbers on the left stand for $N,N'$. 
The cross section $\sigma_{N,N'}$ with $N\geq 6$ are convoluted with a Gaussian of width $1$\ meV.
A few perturbers are indicated by ``PT''.}
\label{deltaN_1}
\end{figure}

\subsubsection{The chain of similar partial cross sections with dominant
configurations only}
Suppose we excite from the ground state only the configurations
$[0\,(N-2)\,1]^+$,  $N =2,3,\ldots$, {\it i.e.} all the principal series. This yields
partial cross sections $\tilde\sigma_{N,N'}$ which already show the
main features of the physical cross sections, and consequently, their
similarity patterns.
The most efficient autoionization mechanism is
governed by propensity rule (A), {\it i.e.},  by reducing the quantum number
$N_{2}$.  Hence, the dipole excited intermediate
configuration $[0\,(N-2)\,1]^+$ will lead to a dominant final
configuration $[0\,(N'-2)\,1]^+$ for the partial cross section $\tilde\sigma_{N,N'}$
with the change in $N_{2}$ being $\Delta N = N-N'$.
The idea is now that partial cross sections 
in different manifolds look similar if their
final 
configurations have an identical
difference $\Delta N = N-N'$ in the quantum number $N_{2}$ with
respect to the respective intermediate configurations.  Fig.~\ref{deltaN_1} shows 
the  partial cross sections across the manifolds $N$ with 
$\Delta N = 1$. Apart from the first cross section $\sigma_{2,1}$ all 
patterns look fairly similar as predicted. There are local 
perturbations marked as ``PT'' and one also notes that the 
similarities become weaker for the highest cross section shown, namely 
$\sigma_{9,8}$. Both of these anomalies we will discuss later, after 
we have explained why $\sigma_{2,1}$ looks so different. This is easy 
to understand because the principal intermediate configuration 
$[0\,0\,1]^+$ in the 
$N=2$ manifold cannot decay through $\Delta N_{2}$ to the $N=1$ 
manifold since $N_{2}=0$ to begin with. Rather, $[0\,0\,1]^+$ decays by 
changing $A$ and $m$ to $[0\,0\,0]^-$ which is not the preferred decay 
route.

Therefore, a chain of similar cross sections has a lower end defined 
by the $(N_1,m)$ quantum numbers of the contributing chain of configurations and its difference 
$\Delta N_{2}= N-N'$ in $N_{2}$:
\be{Nmin}
N_{{\rm min}} = N_{1}+m+\Delta N_{2} + 1\,.
\end{equation}
In our example with $N_{1} = 0$ and $m=\Delta N_{2} = 1$ we have 
$N_{{\rm min}}= 3$, therefore, $\sigma_{2,1}$ does not belong to the 
chain.

\subsubsection{The partial cross section chain including all configurations populated}

\begin{figure}[b]
\epsfxsize  .8\columnwidth
\epsffile{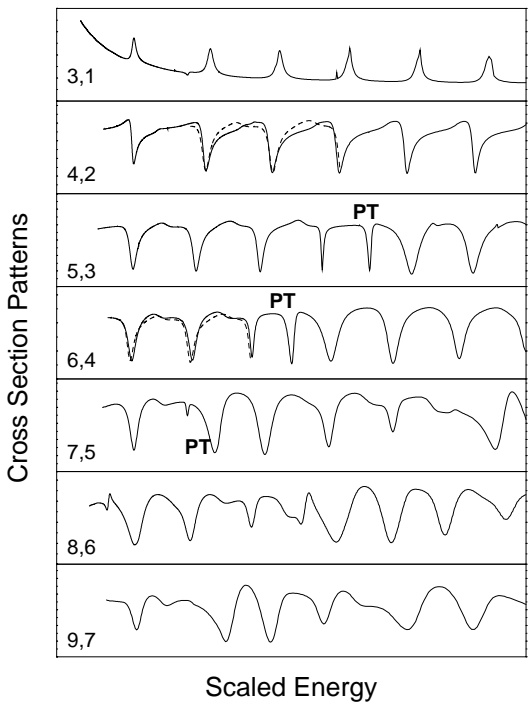}
\caption[]{Same as \fig{deltaN_1} but for the  $\Delta N=2$
partial cross sections. For comparison a clipping of
$\sigma_{N=5,N'=3}$ (dashed line) is shown in the 4,2 and 6,4 panel.}
\label{deltaN_2}
\end{figure}
A closer look on \fig{deltaN_1} reveals that there is still a small 
change in the characteristic pattern from $\sigma_{3,2}$ to 
$\sigma_{5,4}$. The reason is that in addition to the dominant 
intermediate configuration $[0\,(N-2)\,1]^{+}$ other configurations are 
populated as well, each of them having its own chain of similar 
patterns across the manifolds. The actual experimental pattern is the 
sum of all these patterns. However, each chain has its individual 
lower end according to \eq{Nmin}. 
For instance the chain fed by the intermediate configuration 
$[1\,(N-3)\,1]^+$  starts in the manifold $N_{{\rm min}}=4$ and does not 
contribute to $\sigma_{3,2}$. In  fact excited from the ground state 
in helium $[0\,(N-2)\,1]^{+}$ and $[1\,(N-3)\,1]^{+}$ are the two strongest 
intermediate configurations and we expect their chains to be 
sufficient to understand the evolution of the regularity of the 
patterns in the partial cross sections which will be discussed in the 
next section.  For simplicity we introduce a short notation 
\be{chains}
{\cal C}^A_{N_{1},m}(\Delta N_{2})
\end{equation}
to describe the chains where $N_{1},m,A$ characterize the 
intermediate configuration and determine the lower end 
of the chain $N_{{\rm min}}$ according to \eq{Nmin} while $\Delta {N_{2}}$ 
characterizes the type of similar cross sections with $\Delta N = \Delta N_2$
emerging from the chains.  So far we have 
focused on $\Delta {N_{2}} = 1$ (shown in \fig{deltaN_1}) with the two 
dominant chains ${\cal C}^+_{0,1}(1)$ and ${\cal C}^+_{1,1}(1)$.

\section{Systematics in the partial cross sections across the
manifolds from $N=2$ to $N=9$}
\label{sec:pattern}

We will now test the  systematics for the patterns described and 
illustrated in the 
last section  for $\Delta N_{2} = 1$ with  cross sections of $\Delta N_{2} > 1$.
Thereby, we will also discuss the phenomenon of perturbers and the slowly disappearance of 
the patterns for very high partial cross sections, as 
mentioned in the last section.

\subsection{Partial cross sections with $\Delta N = 2$}

We first discuss the $\Delta N = 2$ partial cross sections shown 
in \fig{deltaN_2}.  The general pattern looks quite different 
compared to $\Delta N = 1$ shown in \fig{deltaN_1}. However, 
among each other, the partial cross sections behave similarly as in
\fig{deltaN_1}: The lowest curve $\sigma_{3,1}$ does not match at 
all the other curves, the next one, $\sigma_{4,2}$, is still slightly 
different from the higher ones which are quite similar. For 
$\sigma_{8,6}$ and higher the patterns begin to fade out.
We first note that the lowest possible partial cross section for 
$\Delta N = 2$ is $\sigma_{3,1}$. As for $\sigma_{2,1}$ in
\fig{deltaN_1} the dominant intermediate configuration $[0\,1\,1]^{+}$ cannot decay 
according to the preferred propensity rule $\Delta N_{2} = 2$ but  
must decrease the quantum number $m$ and therefore change the 
quantum number $A$ from $+1$ to $-1$ in addition. Hence, $\sigma_{3,1}$ 
does not belong to the chain of similar cross sections. To the next 
higher one, $\sigma_{4,2}$ contributes only the chain
built on the principal intermediate 
configuration $[0\,2\,1]^+$ with $N_{{\rm min}} = 4$.
For $\sigma_{5,3}$ both dominant chains ${\cal 
C}^{+}_{0,1}(2)$ and ${\cal 
C}^{+}_{1,1}(2)$  can contribute. Consequently, $\sigma_{5,3}$ is the 
partial cross section with the lowest $N$ exhibiting the fully developed pattern of $\Delta N = 2$
which one sees comparing the clipping of $\sigma_{5,3}$ to $\sigma_{4,2}$ and $\sigma_{6,4}$ (see dashed lines in \fig{deltaN_2}).

\subsection{Partial cross sections with $\Delta N = 3$}

\begin{figure}
\epsfxsize  .8\columnwidth
\epsffile{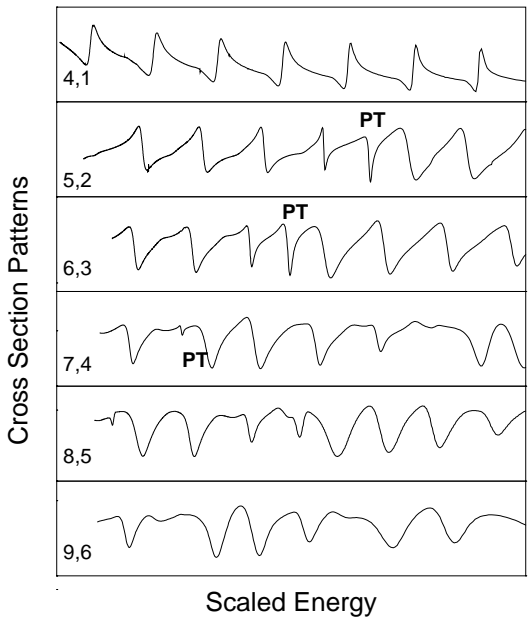}
\caption[]{Same as \fig{deltaN_1} but for the  $\Delta N=3$
partial cross sections.}
\label{deltaN_3}
\end{figure}

These cross sections, shown in \fig{deltaN_3}, have a 
characteristic pattern, which is different from the respective groups 
characterized by $\Delta N =1$ and $\Delta N =2$. Yet, the systematics 
within the group is again the same as for the other two groups and 
can be translated by simply increasing $N$ by one: The first cross 
section $\sigma_{4,1}$  looks extremely different since it does not 
belong to a chain. To the next one only the chain from the principal 
intermediate configuration contributes, $\sigma_{6,3}$ contains for 
the first time the characteristic pattern for $\Delta N = 3$. 
However, since we are already close to $N =8$ where the patterns start 
to fade out due to a beginning break down of the propensity rules to 
which we ascribe their existence, we see only two relatively similar 
cross sections, $\sigma_{6,3}$ and $\sigma_{7,4}$.

\begin{figure}[b]
\epsfxsize \columnwidth
\epsffile{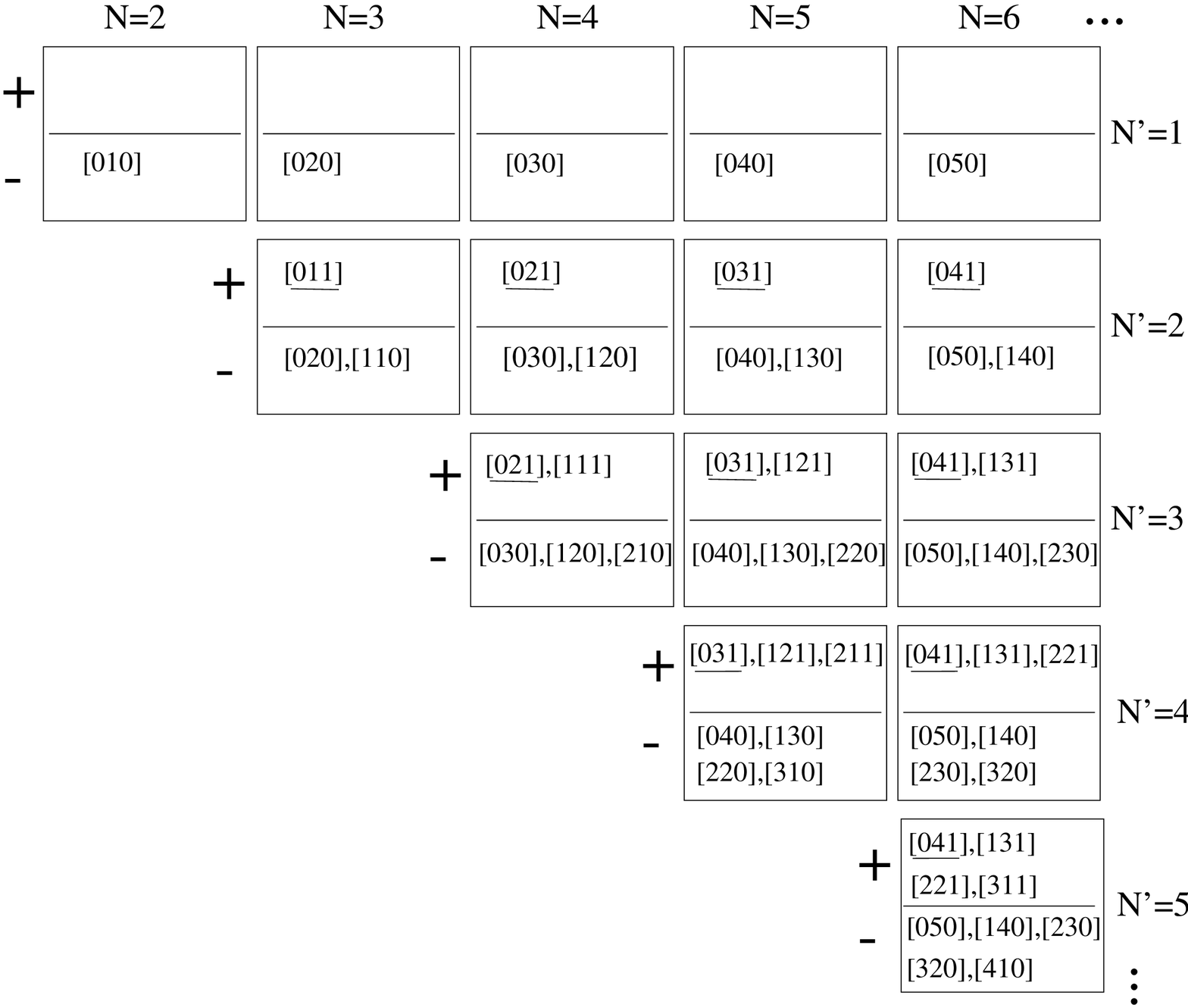}
\caption[]{Compilation of all two-electron configurations (labeled by parabolic
quantum numbers) which can decay according to propensity rule
(A) [\eq{propAA}]. The $+$ and the $-$ signs stand for
$A=\pm 1$ and  the principal configurations are underlined.}
\label{scheme}
\end{figure}

We summarize the systematics of the chains in 
\fig{scheme} where all intermediate configurations are shown
 which can decay according to
propensity rule (A) [\eq{propAA}]. From \fig{scheme} the lower end
characterized by $N,N'$ of any chain can easily be determined. For example,
 ${\cal C}^-_{0,0}(\Delta N)$ evokes a pattern with intermediate configurations
$[0\,N\,0]^-$ already starting at $N'=1$ with  $\sigma_{N,N'=1}$
cross sections. However, as repeatedly pointed out, they are too weak
to be seen in the cross sections.

\subsubsection{The role of isolated perturbers}
Before the pattern actually breaks down 
(see, e.g., \fig{deltaN_3})
it can  already be locally 
distorted by so called perturber states.
As is well known from quantum defect theory \cite{WiFr}, a perturber
acts in a twofold way on the resonance states to which it couples.
Firstly, it shifts their positions (``bunching effect'') which is
expressed by a jump in the quantum defect. Secondly, it modulates their
linewidth in a Fano-profile like way. Both effects locally perturb
the cross section pattern.

\begin{figure}
\epsfxsize  \columnwidth
\epsffile{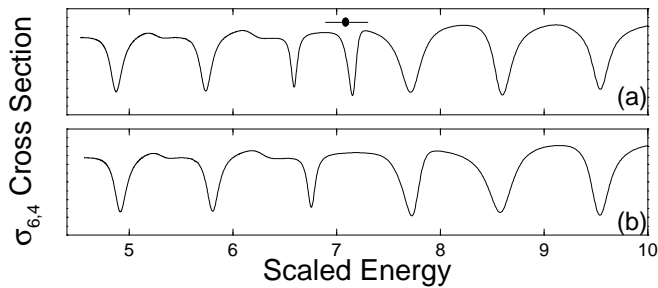}
\caption[]{Partial cross section of the $N'=4$ satellite below the
$N=6$ threshold (a) as a function of the effective quantum number
$\nu_6(E)$ [\eq{scaling}] and (b) as a function of
$\nu_6(E)+\delta_{6,4_n}(E)$, where $\delta_{6,4_n}(E)$ denotes the
quantum defect of the series $6,4_n\equiv [0\,4\,1]_n^+$. In (a) the
position and the linewidth of the perturber $N,K_n=7,5_7$ are
indicated \cite{Roal97}. In (b) the bunching of the resonances due to
the perturber is disentangled restoring the similarity pattern.}
\label{unfold}
\end{figure}

The bunching of the resonances is visible in the cross
sections $\sigma_{N=5,N'}$ and $\sigma_{N=6,N'}$ with the perturbers
$N,K_n=6,4_6$ and $7,5_7$, respectively.  The distorting influence on
the pattern can be compensated by incorporating the quantum defect
$\delta_{N,K_n}(E)$ of the perturbed series.  Plotting the cross
section against the effective quantum number
$\nu_6(E)+\delta_{6,4_n}(E)$, restores the characteristic pattern of
the cross section as can be seen in \fig{unfold}(b).  The effective
quantum defect compensates the bunching of resonances on the energy
axis.  Therefore, this kind of disentanglement works well, as long as
the effect of the perturber on the width of the resonances is small as
it is the case for the cross sections $\sigma_{N=5,N'}$ and
$\sigma_{N=6,N'}$.  The perturber $8,6_8$, however, causes a drastic
narrowing of the width of the state $7,5_{10}$.  This perturbation of the
pattern cannot be compensated by expressing the energy in terms of
the effective quantum defect.  However, the perturbation  remains small 
and local leaving the general pattern still identifiable
as on can see in the cross sections $\sigma_{7,N'}$ of
\figs{deltaN_1}{deltaN_2}{deltaN_3}.

\subsubsection{Fading out of the patterns}
Going to higher manifolds the  patterns start to fade out.
This is certainly due to an increasing number of perturbers.
However, in more general terms, this observation indicates the beginning
break-down of approximate quantum numbers and consequently of propensity rules
which govern the patterns. This refers to a situation discussed here 
with the principal quantum number $n$ of the outer electron only 
moderately larger than  the principal quantum number $N$ of the inner 
electron. Clearly, for $n\gg N$ the regime of a (regular)  effective
one-electron Rydberg series is always approached.
\section{Conclusions}

We have presented numerical total and partial cross sections
for single photoionization from the helium ground state up to the
$N=9$ threshold of \hp. Our calculations were done by using the
eigenchannel R-matrix method. We found very good agreement with
available experimental data for both the total cross section up to
the $N=9$ manifold and the partial cross sections up to the $N=5$
manifold.

A comparison of the partial cross sections $\sigma_{N,N'}$ ($N'$
denoting the state of the residual helium ion) across the manifolds
reveals common patterns in the cross sections with the same $\Delta
N=N-N'$.  The patterns of the principal series dominates with a
seizable contribution from strongest secondary series due to the large
oscillator strength of these series.  The manifestation of the
patterns can be attributed to chains of configurations which connect
the intermediate configurations of resonance states seen in the cross
sections to final configurations in the different continua according
to the dominant propensity rule for autoionization.

Starting with the $N=5$ manifold perturbers emerge which locally
destroy the general patterns.  However, in cases where the 
perturber mainly leads to a bunching of resonances on the energy axis, a
regularization based on energy-dependent quantum defects has been shown
to disentangle the spectra and restore the similarity of the 
patterns. Going to manifolds $N = 8$ and higher the patterns start to fade
out which finally indicates the break-down of the propensity rules.
This in turn signals the approaching limits of the adiabatic picture
and the approximate quantum numbers derived from it.


\begin{acknowledgments}
We would like to thank A. Menzel and R. P{\"u}ttner for providing us
with their experimental data. Financial support by the
DFG  through the Gerhard Hess-program is gratefully acknowledged.
\end{acknowledgments}


\end{document}